\documentclass[10pt,letterpaper]{article}
\usepackage{opex3,amsmath}

\newcommand{\comment}[1]{}
	
\begin{document}	

\title{Using artificial neural networks for open-loop tomography}
              
\author{James Osborn,$^{1*}$ Francisco Javier De Cos Juez,$^2$ Dani Guzman,$^1$ Timothy Butterley,$^3$ Richard Myers,$^3$ Andr\'{e}s Guesalaga,$^1$ and Jesus Laine$^2$}	 
\address{$^1$Dept. of Electrical Engineering, Centre for Astro-Engineering, Pontificia Universidad Catolica de Chile, Vicu\~{n}a Mackenna 4860, Santiago, Chile\\
$^2$Mining Exploitation and Prospecting Department, C/Independencia n¼13, University of Oviedo, 33004 Oviedo, Spain \\
$^3$Department of Physics, Centre for Advanced Instrumentation, University of Durham, South Road, Durham, UK, DH1 3LE}

\email{* josborn@ing.puc.cl}
\date{\today}					

\begin{abstract}	
Modern adaptive optics (AO) systems for large telescopes require tomographic techniques to reconstruct the phase aberrations induced by the turbulent atmosphere along a line of sight to a target which is angularly separated from the guide sources that are used to sample the atmosphere. Multi-object adaptive optics (MOAO) is one such technique. Here, we present a method which uses an artificial neural network (ANN) to reconstruct the target phase given off-axis references sources. We compare our ANN method with a standard least squares type matrix multiplication method and to the learn and apply method developed for the CANARY MOAO instrument. The ANN is trained with a large range of possible turbulent layer positions and therefore does not require any input of the optical turbulence profile. It is therefore less susceptible to changing conditions than some existing methods. We also exploit the non-linear response of the ANN to make it more robust to noisy centroid measurements than other linear techniques.
\end{abstract}

\bibliographystyle{osajnl}

\section{Introduction}

Adaptive optics (AO) systems require guide sources to sample the turbulent atmosphere above the telescope. If a guide star is located very close to the target or we can use the target itself, then this star can be used to directly measure the phase aberrations along the line of sight to the target. However, if there is no guide star bright enough or we would like to observe multiple or extended objects in the field then we require multiple guide stars to sample the turbulent atmosphere. Another reason for using multiple guide sources is when artificial guide stars are employed. In this case each guide star only illuminates a cone within the turbulent volume above the telescope. If the light cones of these guide stars overlap with the cylinder illuminated by the target we can use tomographic techniques to reconstruct the phase aberrations along the line of sight to the target. The majority of modern AO systems (with the exception of extreme AO for extrasolar planet imaging) make use of tomographic reconstruction techniques. Three major varieties of tomographic AO currently under investigation are laser tomography AO (LTAO) \cite{Louarn04}, multi-conjugate AO (MCAO) \cite{Beckers89} and multi-object AO (MOAO) \cite{Hammer02,Assemat07}. 

In the case of MOAO a number of target directions are observed simultaneously and corrected independently by one deformable mirror (DM) per channel. The guide stars (natural and laser) are distributed around the field and are monitored with open loop wavefront sensors (WFS) (i.e. without a DM). The information from each guide star is then combined in such a way as to estimate the phase aberrations for each target. CANARY \cite{Morris10,Gendron11} is the first on-sky test of tomographic MOAO and is thus a perfect test bench for both the opto-mechanical technology that needs to be developed and for the algorithms that are required for the control of the instrument.\comment{ For this reason we plan to test the reconstructor with CANARY and the parameters for our simulations have been chosen to mirror those of CANARY.}

Optical turbulence profilers show that the atmosphere can be considered to be made up of a number of independent very thin turbulent layers. The altitude and strength of these layers can change with time and so the vertical profile of the optical turbulence will develop and evolve with time \cite{Avila06}. Median profiles are used in simulations for performance analysis but it should be remembered that a median profile is not representative of any real profile. The tomographic reconstructor must be able to handle these changes of turbulent profiles.

Here we present a new method which uses an Artificial Neural Network (ANN) to combine the information from the WFSs and output the integrated reconstructed phase aberrations from the target to the telescope. ANNs are trained by exposing them to a large number of inputs together with the desired output. In theory this training data should cover the full range of possible scenarios. However, this is obviously not possible and so given enough training the ANN will provide a best guess to the solution. When the ANN is confronted with a superposition of a number of the independent training sets it can then predict an output by combining a number of the synaptic pathways. In this way we do not need to train the ANN with every possible turbulent profile.

We propose to train an ANN off-line with simulated data. The reconstructor is named CARMEN (Complex Atmospheric Reconstructor based on Machine lEarNing). The idea is to train the reconstructor to be able to handle any turbulent profile that it might be exposed to. We do this by carefully selecting the optimum training routines. This is a train and apply technique; once trained with the correct parameters (for example, the number and geometry of the guide stars) it will work for any optical turbulence profile. Therefore, we train CARMEN with a large number of independent turbulence profiles. We train it with the off-axis WFS slopes and the desired on-axis target Zernike coefficients. When the network is implemented and shown the off-axis WFS data it will estimate what the on-axis Zernike coefficients will be. We have chosen to output Zernike coefficients at this stage to limit the number of outputs required (i.e 27 values, assuming we predict up to $6^{\mathrm{th}}$ order Zernikes, instead of a value of the order of the number of actuators in the DM). It would be possible to predict a higher number of degrees of freedom and we would expect the performance to increase accordingly. No {\it a priori} knowledge of the atmosphere is required and no input from the user or re-training is required if the atmospheric turbulence profile changes during observing. This is an alternative approach to most other tomographic reconstructors.

ANNs have been applied to the field of AO in the past. Angel {\it et al.} (1990) \cite{Angel90}, Sandler {\it et al.} (1991) \cite{Sandler91} and Lloyd-Hart {\it et al.} (1992) \cite{Lloyd-Hart92} present successful results using neural networks for wavefront sensing in the focal plane. Montera {\it et al.} (1996) \cite{Montera96} experimented with an ANN to reduce WFS centroiding error and to estimate the Fried parameter $r_{0}$ and the WFS slope measurement error. They found that the ANN performed as well as but not better than a standard linear approach for estimating the WFS slopes and for the estimation of the Fried parameter, r0, however the ANN was very good at estimating the WFS slope measurement error. ANNs have also been investigated for spatial and temporal predictions of the slope measurements. Lloyd-Hart \& McGuire (1995) \cite{Lloyd-Hart95} use an ANN to make a temporal prediction of the WFS slopes. The AO latency is then reduced allowing for a better correction. Weddell \& Webb (2006, 2007) \cite{Weddell06, Weddell07} developed this idea and used off-axis WFS measurements to temporally predict the on-axis slopes. However, this was limited to low-order Zernike modes (tip/tilt) only. More recently neural networks have  been used to model open loop DMs for MOAO \cite{Guzman10}. An accurate DM model is required for open-loop AO as the DM is not seen by the WFS.

The difference between our proposal and the work of Lloyd-Hart \& McGuire (1995) \cite{Lloyd-Hart95} and Weddell \& Webb (2006) \cite{Weddell06} is that we will train the network in simulation rather than on-sky. This allows us to select and control what the network learns and means that we can predict to a higher order. One advantage of the on-sky training is that it will inherently be trained to the concurrent turbulence profile. However, if this profile were to change then, like other reconstructors that need to be re-calcualted, it would need to be re-trained.

In section 2 we discuss two of the existing tomographic reconstructors. These are the least squares matrix vector multiplication (LS) and learn and apply (L+A) techniques. We will use these two reconstructors as a benchmark to test CARMEN. In section 3 we describe the neural network, present the optimum training scenario used and briefly state other approaches which were investigated. In section 4 we present the tomographic results of CARMEN and the two other reconstructors from simulation of three test atmospheric profiles. We also show the effects of photon and read noise in the wavefront sensor. In section 5 we discuss some of the issues related to implementing CARMEN in a real AO system.

\section{Existing reconstructor techniques}

\begin{figure}[hbt]
   \centering
   \includegraphics[width=51mm]{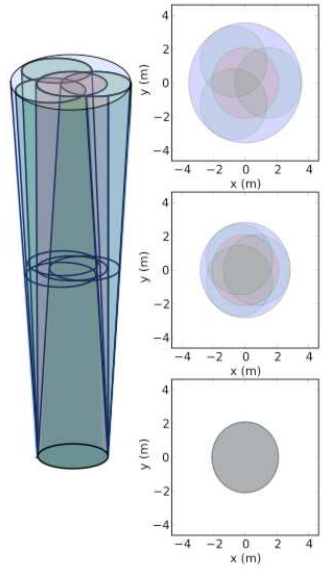}
   \caption{Topological diagram of the light cones for three guide stars and one target for a 4.2~m telescope and guide stars equally distributed on a ring of radius 30 arcseonds. The target direction is shown in red, the guide stars in green and the full field of view in blue. The cut-throughs on the right are taken at 0~m, 5000~m and 10000~m. At higher altitudes the overlap of the guide stars reduces and we sample smaller areas of the target light cone.}
   \label{fig:overlap}
\end{figure}

Tomographic reconstruction is the re-combination of the information from several guide stars to estimate the phase aberrations along a different line of sight to a scientific target. A standard approach is to use Shack-Hartmann WFSs to measure the phase aberrations in the light cones to the guide stars. When the light cones overlap at the altitude of a turbulent layer the same phase aberrations will be applied to both wavefronts but in different areas of the meta-pupil. We can then look for correlation in the phase maps at the ground. Figure~\ref{fig:overlap} shows a topological diagram of a system with three guide stars and one target. Any turbulence at low altitudes will be well sampled. At higher altitudes the overlap is reduced and we therefore have less information. Above the altitude where the beams no longer overlap there will be very limited correlation in the phase aberration (possibly some correlation in the very low order modes, depending on the extent of the separation and the outer scale of the turbulence) and it is therefore very difficult to gain any information. Any turbulence above this altitude will essentially be noise.

There are several tomographic techniques which can be used to combine the information from the guide stars. Here we examine two methods, a standard least squares type matrix vector multiplication (LS) and learn and apply (L+A). We have chosen these two as benchmark tests to compare with our new technique.

The standard LS method (e.g \cite{Ellerbroek94,Fusco01}) involves multiplying the WFS vectors with a control matrix. The control matrix maps the response of the wavefront sensors to the actuator commands of the DM(s) and can be computed off-sky. The technique is computationally intensive. Although this is not a problem for current telescopes and AO systems, for the next generation of extremely large telescopes and modern AO systems, which are either high order or include a number of wavefront sensors and DMs, the LS method may become unfeasible. There has been a lot of interest in sparse LS solutions to alleviate this problem \cite{Wild95,Thiebaut10}. In the case of MOAO we place `virtual' DMs at the conjugate altitude of the turbulent layers in the atmosphere and calculate the control matrix of each of these independently by either telescope simulator on an optical bench or simulation. It is very important to position these virtual DMs accurately at the conjugate altitude of the turbulent layers or the performance will be compromised. If the profile of optical turbulence was to change during observation the tomographic reconstructor would provide a poor fit to the actual slopes. We therefore require high vertical resolution atmospheric optical turbulence profiles in order to optimise this tomographic reconstructor.

Learn and Apply (L+A) \cite{Vidal10} has recently been developed and successfully tested with CANARY. L+A has taken a different approach to many other techniques in that it includes the concept of a SLODAR \cite{Wilson02} system and so automatically includes the atmospheric optical profile within the reconstruction. This is done by calculating the covariance matrices between the slopes of all of the guide stars with each other and all of the guide stars with an on-axis calibration WFS. By combining the two covariance matrices, the turbulence profile and geometric positions of all the guide stars with the target are taken into account in the reconstructor. If the turbulence profile were to change during the course of the observation the covariance matrices would need to be re-calculated. However, as the guide star WFS are open loop it is possible to monitor the profile using the SLODAR method and therefore know when the reconstructor needs to be updated. It should be noted that the on-axis WFS is only available during calibration and not when the instrument is observing the scientific target. This means that the reconstructor can not be completely updated during observations. However, it might be possible to estimate the on-axis covariance matrix using the off axis matrix and knowledge of the geometry, allowing L+A to be stable even in changeable conditions.

\section{Neural networks}
\begin{figure}[htb]
   \centering
   \includegraphics[width=70mm]{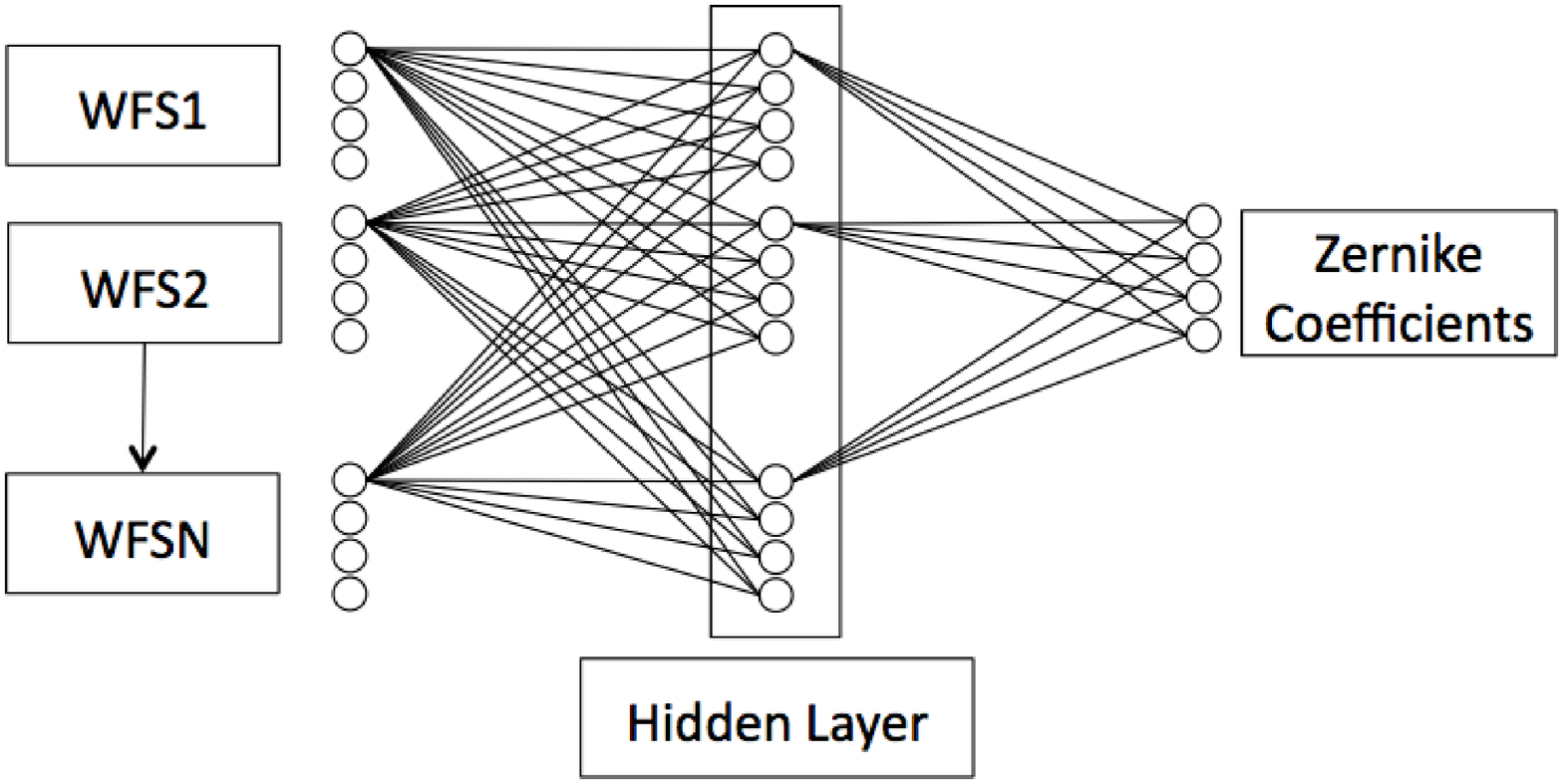}
   \caption{A simplified network diagram for CARMEN. The slopes from the WFS are input to the network. They are all connected to every neuron in the hidden layer by a synapse. Each neuron in the hidden layer is then connected to every output node. CARMEN will output the predicted Zernike coefficients for the target direction. Each of the synapses has a weight. At run time the inputs are injected into the network which is then processed by the different activation functions and weights generating a response. In the diagram only a few of the synapses are shown for clarity.}
   \label{fig:neural_net}
\end{figure}

ANNs are well known for their ability to solve problems that are otherwise difficult to model \cite{Huarng06,Swingler96,Denton95}. A detailed explanation of ANNs can be found in \cite{Guzman10}. Artificial Neural Networks are computational models inspired by biological neural networks which consist of a series of interconnected simple processing elements called neurons or nodes. The Multi Layer Perceptron (MLP) is a specific type of Feedforward Neural Network in which the nodes are organized in layers (input, hidden and output layers) and each neuron is connected with one or more of the following layers only. Each neuron receives a series of data from the preceding layer neurons or an external source, transforms it locally using an activation or transfer function (equation~\ref{eqn:1}) and sends the result to one or more nodes in any of the following layers (figure~\ref{fig:neural_net}). This cycle repeats until the output neurons are reached.
 
Each connection between neurons has a numerical value which represents the importance of the preceding neuron\comment{ in the result of the actual one}, called ``synaptic weight'', $w$. It is in these values where the most important fraction of information is stored \cite{Haykin99}. There is also a special type of neuron called ÒbiasÓ which has no connection with neurons in the previous layers, but can apply to every neuron in a given layer. It enhances the flexibility of the network. Mathematically the output of the $j^{\mathrm{th}}$ neuron, $y_{j}$, is computed as,
\begin{equation}
y_{j}(t) = g(x_i)=g\left(\sum_{i} w_{ji} \cdot y_i+b_j\right),
\label{eqn:1}
\end{equation}
where $g(\cdot)$ is the activation function which is applied to the input of the neuron, $x_{i}$. This local input is the sum of all the output values, $y_{i}$, sent by the feeding neurons from the previous layer, multiplied by the corresponding synaptic weight, $w_{ji}$\comment{, which connects the $i^{\mathrm{th}}$ neuron with the $j^{\mathrm{th}}$}, plus the bias weighted value, $b_j$. $i$ is an index that represents all of the neurons which are feeding the $j^{\mathrm{th}}$ neuron.

The network needs to be trained before it can be used. During\comment{ the} training, the weights are changed to adopt the structure of a determined function, based on a series of input-output data sets provided. The Backpropagation (BP) training algorithm, used in this work, attempts to minimize the least mean square difference over the entire training set \cite{Rumelhart86}. When the network is set up, the connections between the neurons\comment{ in the input and hidden layers} are assigned random weights, the network then produces an output using equation~\ref{eqn:1}. The least mean square difference is then computed as,
\begin{equation}
E=\frac{1}{2}\sum_k\sum_m \left( y_{km}-d_{km}\right)^2,
\label{eqn:2}
\end{equation}
where $y_{km}$ is the net predicted value, $d_{km}$ is the target, $k$ goes from 1 to the number of input-output sets and $m$ is an index that represents each output node. Usually the weights increase proportionally to the negative gradient of the least mean square error,
\begin{equation}
\Delta w_{ij} = -\epsilon \frac{\partial E}{\partial w_{ij}} = -\epsilon \frac{\partial E}{\partial x_{i}}\cdot \frac{\partial x_i}{\partial w_{ij}} = -\epsilon \frac{\partial E}{\partial x_{i}}\cdot \frac{\partial }{\partial w_{ij}} \sum_j w_{ij}\cdot y_j = -\epsilon\cdot \delta_i \cdot y_j.
\label{eqn:3}
\end{equation}
The proportional constant $\epsilon$ is called the Òlearning rateÓ and is a key parameter of neural network training \cite{Haykin99, Rumelhart86}. So, we compute all errors ($\delta_i$) for all the neurons involved in the net. The algorithm is able to do this by first computing the state for all the neurons (hidden included) using equation~\ref{eqn:3}, and then computing the error of all the nodes by means of the states, errors and weights of the next layers by applying the chain rule,
\begin{equation}
\delta_i = \frac{\partial E}{\partial x_j} = -\sum_i \frac{\partial E}{\partial x_i}\cdot \frac{\partial x_i}{\partial y_j}\cdot \frac{\partial y_j}{\partial x_j}=g^{\prime}(x_j)\cdot\sum_i \delta_i\cdot w_{ij}.
\label{eqn:4}
\end{equation}
This process of computing weight changes for all the data repeats a number of times defined by the user or until an error value is reached. Each iteration is called an ÒEpochÓ.
In general, the more complex the problem, the smaller the permitted adjustment each epoch. Otherwise the network will not detect subtle patterns within the data, and as the ANN sequentially overcompensates, the error curve will oscillate wildly rather than converging to the global minima.  It is important that the training data contain adequate numbers of Òpossible atmospheric casesÓ if the ANN is to be used for a prediction problem. Over-fitting can be a problem if the data sample is too small, biased, or the network has too many nodes. In essence, the network sacrifices the ability to generalize in order to achieve the most accurate fit to the training data \cite{Rumelhart86}. Once developed and trained on retrospective data, the ANN must, as with all statistical models, be validated by previously unseen data from a different data set \cite{Bottaci97}. \comment{The final essential step involves validation on prospective data.}

\subsection{Training}
As described above, the ANN is trained by showing it a representative selection of inputs with the desired outputs. The training data should attempt to cover the full range of possible scenarios. We propose to train the ANN with simulated data. If we present it with enough independent data the weightings will converge and the network should be able to cope with any input which is similar to, or a combination of stimuli which are all similar to, the training data. If we are not careful with the training data the network will learn to make connections which are only a coincidence in the training set or are perhaps a secondary concern. By using simulated data we can control what the neural network sees and hope to guide the learning process. We have tested many training scenarios. The best one we have found involves training the network with a single turbulent layer ($r_0=0.12$~m and $L_0=30$~m) . The layer is placed at 155 altitudes ranging from 0 m to 15500 m with 150 m resolution. At each altitude we present CARMEN with 1000 randomly generated phase screens. Using this dataset CARMEN has seen all of the possible layer positions. CARMEN will combine the response of this basis set and use it to model the input data. We can essentially model the atmosphere with the same resolution we use to train CARMEN. In reality what we are doing is teaching the network how to combine slopes with different light cone overlap fractions in the WFSs (figure 1). 
There are other alternatives for training sets, like including two turbulent layers, one fixed at the ground and another higher layer at a number of different altitudes\comment{[8-10]} or a more realistic case with a number of layers with different relative strengths. However, although more realistic, these datasets are no longer independent, the network is over trained and we find that the results are not as good as with the simpler approach explained above. 

After testing networks with different network architectures and actuation functions we have found that the optimum architecture depends on the profile of the optical turbulence in the atmosphere and on the magnitude of the noise. As the optimum architecture is different under different conditions we have decided to use the simplest approach which produces good results in all cases. The simplest network consists in a MLP of only one hidden layer containing the same number of neurons as the input allowing full
mapping, and BP training algorithm with a sigmoid activation function and a value of learning rate of 0.03. The results from the more complicated networks are not presented here, however they were all broadly similar with each one having slightly better performance in different circumstances. For example, in more complicated atmospheres with many turbulent layers networks with an additional hidden layer resulted in a slightly lower residual wavefront error. By training the networks with these simplistic sets that cover the full range of possible layer positions the network can combine the responses in order to estimate the outputs from much more complicated profiles. No additional information or re-training is necessary even if the atmosphere changes drastically during observing. The tomographic reconstructor is robust even in the most challenging conditions.

The hardware used for training was an OpenSuSe 11.3 running on a 8 core 2.4 GHz Intel Xeon CPU E5530 with 32 Gb RAM, although only 1 core and nearly 620 Mb of RAM were used. With this configuration using R v2.12.2 the training time was of 4 days 1 hour and 23 minutes.

\section{Results}

The results presented here are generated by Monte Carlo simulation. We assume three off-axis natural guide stars equally spaced in a ring of 30 arcseconds radius. The target direction is at the centre of this ring. The telescope diameter is 4.2~m and we assume $7\times7$ subapertures in the Shack-Hartmann WFS. The simulation parameters were chosen to be similar to those of CANARY and the results are compared with a standard LS method and with L+A. In the simulations we use a standard thresholded centre of gravity algorithm for the centroiding. 

CARMEN is trained to return the first six radial orders of Zernike coefficients (not including piston) rather than the subaperture slopes. This was done to reduce the computational load during training for a more efficient investigation. However, it should be noted that there is no reason\comment{, given enough time,} why the system could not be trained to return slopes instead. For a fair comparison we apply all of the reconstructors to a modal DM, correcting to the same number of Zernike modes. The reconstructed Zernike phase is subtracted from the pupil phase and then used to generate the point spread function (PSF). The metrics used to asses the results are wavefront error (WFE [nm]), PSF Strehl ratio, azimuthally averaged PSF full-width at half maximum (FWHM [arcseonds]) and diameter of 50\% encircled energy (E50d [arcseconds]) in the H-band (1650~nm). The WFE includes the tomographic error and the fitting error of the six radial orders of Zernikes to the real phase.

We assess each of the tomographic reconstructors with three test cases. These are the good, median and bad seeing atmospheric profiles from La Palma, as used in the CANARY simulations\comment{derived from \cite{Fuensalida07} and are used in the CANARY simulations} (shown in table~1). Each of the profiles have four turbulent layers, but the altitudes and the relative strengths of the layers and the integrated turbulence strength is different in each case.
\begin{table*}   
\begin{minipage}{140mm}
\centering 
\begin{tabular}{|l| c| c| c| c| }  
\hline
\hline
Parameter & \multicolumn{3}{|c|}{Values} & Units \\ [0.5ex] 
\hline
Test Name & atm1 & atm2 & atm3 &\\
\hline
$r_{0}$ (at $0.5\mu m$)      &  0.16& 0.12&0.085  &m\\
\hline   
Layer 1   &    &&    & \\  
Altitude                  & 0         &0         &0   & m\\
Relative strength   &  0.65   &0.45    &0.80  &\\
Wind Speed          & 7.5       &7.5      &10  & m/s\\
Wind direction       & 0          &0         &0    & degrees\\[1ex]   
\hline    
Layer 2   &    &&    & \\  
Altitude                  & 4000   &2500    &6500   & m\\
Relative strength   &  0.15    &0.15     &0.05  &\\
Wind Speed           & 12.5    &12.5     &15  & m/s\\
Wind direction        & 330     &330      &330   & degrees\\[1ex]   
\hline    
Layer 3   &    &&    & \\  
Altitude                  & 10000  &4000    &10000   & m\\
Relative strength   &  0.10    &0.30     &0.10  &\\
Wind Speed           & 15       &15         &17.5  & m/s\\
Wind direction        & 135     &135       &135   & degrees\\[1ex]   
\hline    
Layer 4   &    &&    & \\  
Altitude                  & 15500  &13500 &15500   & m\\
Relative strength   &  0.10    &0.10    &0.05  &\\
Wind Speed          & 20        &20        &25  & m/s\\
Wind direction       & 240      &240     &240   & degrees\\[1ex]   
\hline\hline

\end{tabular} 
\caption{Table of atmospheric parameters for the three test cases. The outer scale is 30~m.}
\end{minipage}
\label{table:atms}  
\end{table*}  

In order to compare the reconstructors fairly the LS method is optimised in terms of virtual DM altitude and actuator density by experimentation and the learn stage of L+A is also performed with an atmosphere of the same parameters, but different phase maps, for each test case.

\subsection{Noiseless simulation results}

The results of a noiseless simulation are shown in table~2. On the contrary to the LS and L+A reconstructors, no change was made to CARMEN between the test cases. The results show that CARMEN was able to adapt to each test case successfully as it consistently results in the lowest WFE. It is important to note that we show the comparison results to prove that CARMEN can perform as well as the other techniques. The other techniques can be optimised to further improve these results, however the real strength of CARMEN is that it does not need any modification even in very changeable conditions and this is reflected in the results. 

If the exposure time was long enough for even the lowest order modes to average out then if we were to run each of these test cases sequentially to simulate a changing atmosphere then the resulting PSF will simply be the sum of the three test case PSFs. Therefore, we see that CARMEN would be able to function with a changing atmosphere with no re-configuration necessary. The other reconstructors would require re-configuring in order to obtain a similar result, otherwise the performance could be seriously impaired.
\begin{table*}   
\begin{minipage}{140mm}
\centering 
\begin{tabular}{|c| c| c| c| c| c|}  
\hline
\hline
Test Name & Reconstructor &\multicolumn{4}{|c|}{Metrics} \\
\hline
& &Strehl ratio & FWHM (arcsec) & E50d (arcsec) & WFE (nm)\\ [0.5ex] 
\hline                    
atm 1 & Uncorrected        & 0.048 & 0.319 & 0.482 & 644\\   
          & LS            & 0.296 & 0.099 & 0.299 & 293\\
          & L+A          & 0.402 & 0.089 & 0.293 & 251 \\
          & CARMEN  & 0.462 & 0.088 & 0.279 & 231 \\  [1ex]
\hline    
atm 2 & Uncorrected        & 0.025 & 0.458 & 0.633 & 817\\   
          & LS            & 0.230 & 0.100 & 0.443 & 322\\
          & L+A            & 0.300 & 0.091 & 0.436 & 289 \\
          & CARMEN  & 0.370 & 0.088 & 0.393 & 262 \\  [1ex]    
\hline    
atm 3 & Uncorrected        & 0.012 & 0.684 & 0.912 & 1088\\   
          & LS            & 0.068 & 0.143 & 0.690 & 454\\
          & L+A            & 0.100 & 0.104 & 0.688 & 409\\
          & CARMEN  & 0.125 & 0.101 & 0.660 & 387\\  [1ex]   
\hline\hline

\end{tabular} 
\caption{Table of PSF metrics for each tomographic reconstructor and test scenario. All metrics, except WFE, are defined at 1650~nm.}
\end{minipage}
\label{table:Results}  
\end{table*}

\begin{figure}
   \centering
   $\begin{array}{cc}
   \includegraphics[width=65mm]{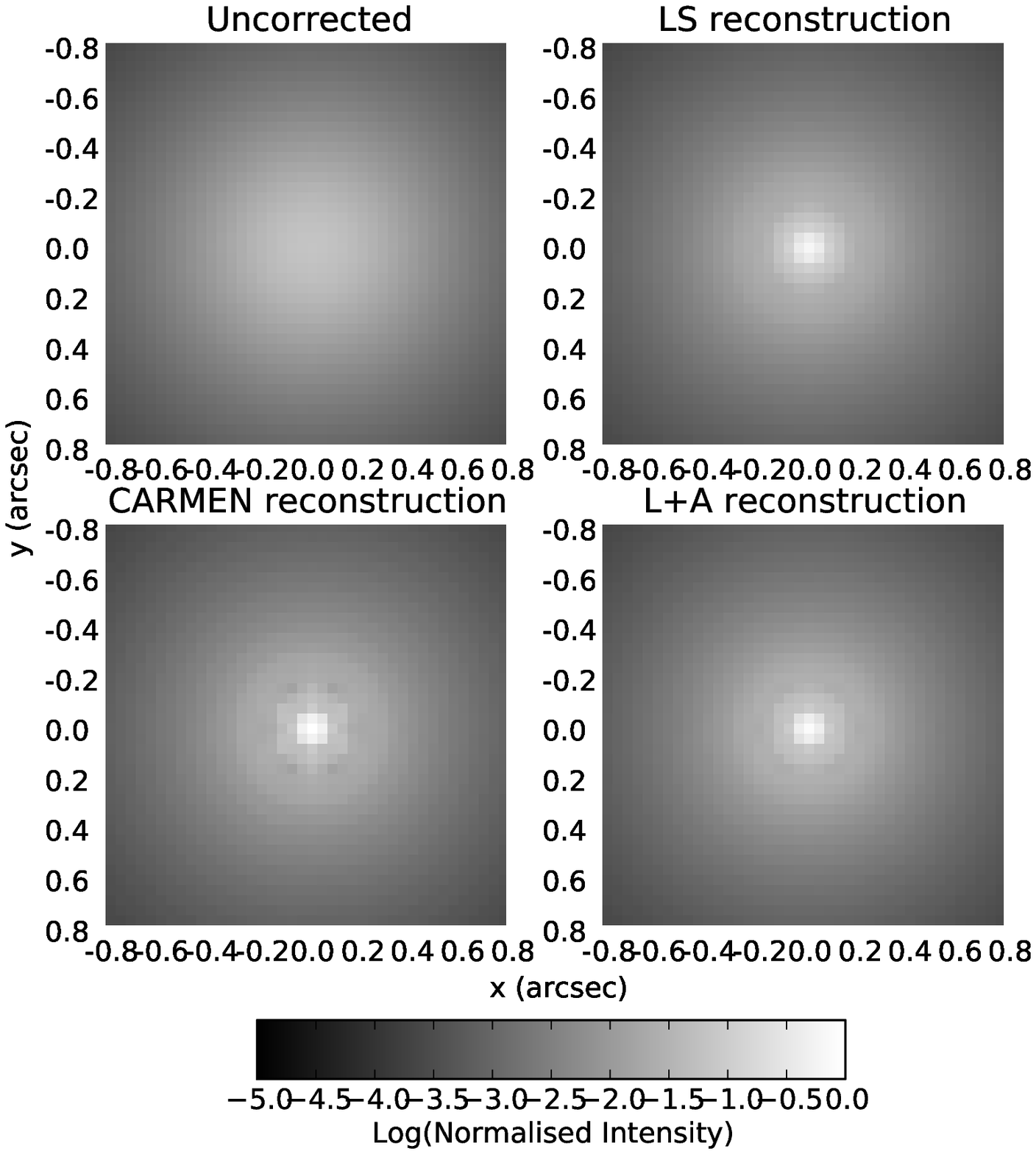}&
   \raisebox{10mm}{\includegraphics[width=65mm]{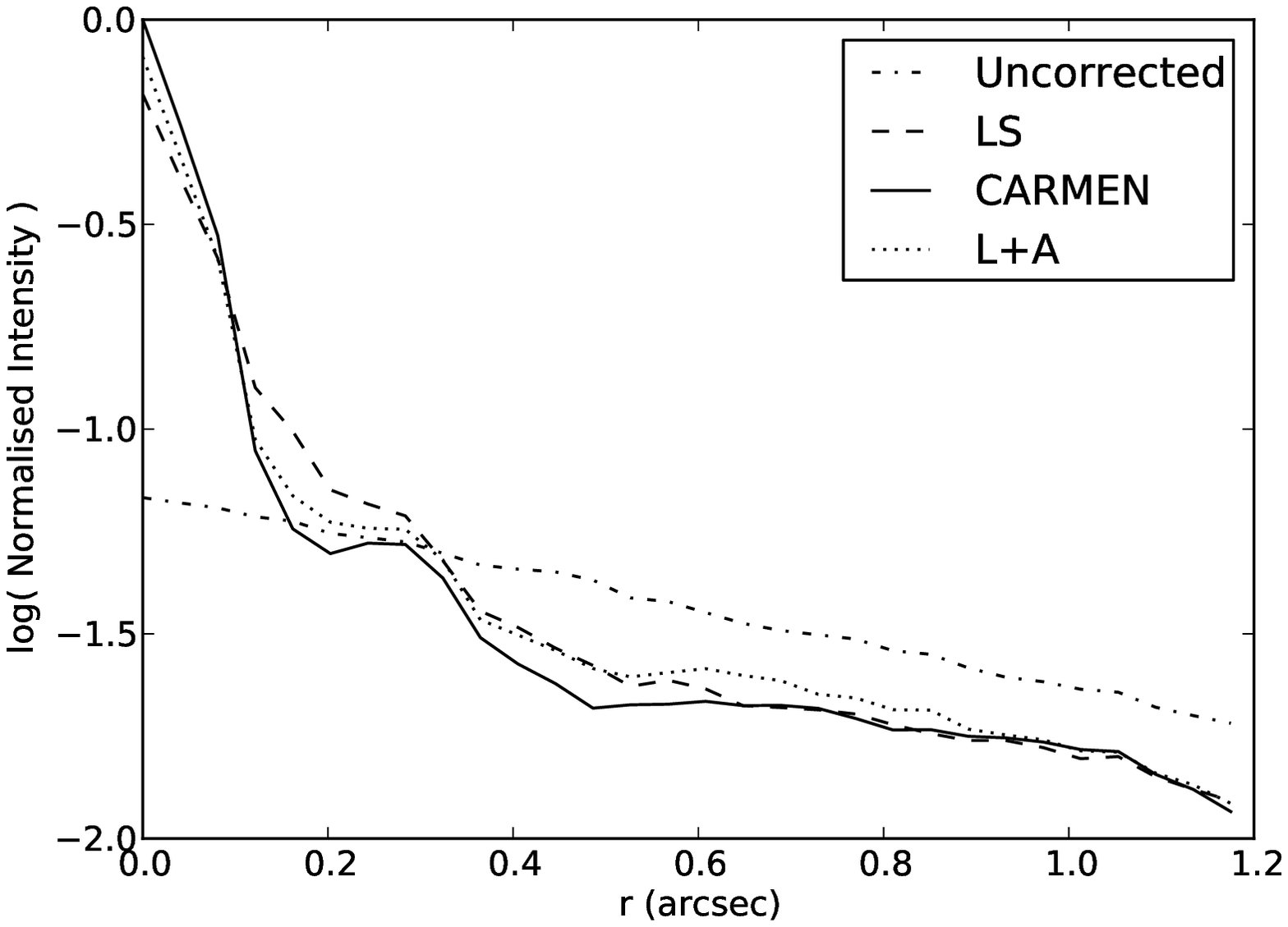}}\\
       \mbox{\bf (a) } & \mbox{\bf (b)} 
      \end{array}$
      \caption{Simulated PSFs (left) for test 2 (median seeing scenario). Clockwise from top left is the uncorrected PSF, LS, L+A and CARMEN tomography. The residual WFEs are 817~nm, 322~nm, 289~nm and 262~nm respectively. The azimuthally averaged radial profiles of the PSFs are shown on the right.}
   \label{fig:test2_rad_prof}
\end{figure}

Figure~\ref{fig:test2_rad_prof} (a) shows the PSFs generated using each of the tomographic reconstructors and the atomospheric test case 2 (median seeing). The azimuthally averaged radial profiles are also shown in figure~\ref{fig:test2_rad_prof} (b). The non-circular diffraction effect seen in the PSF is because we are approximating the wavefront with Zernikes up to sixth order.

\begin{figure}
   \centering
    $\begin{array}{cc}
    \includegraphics[width=65mm]{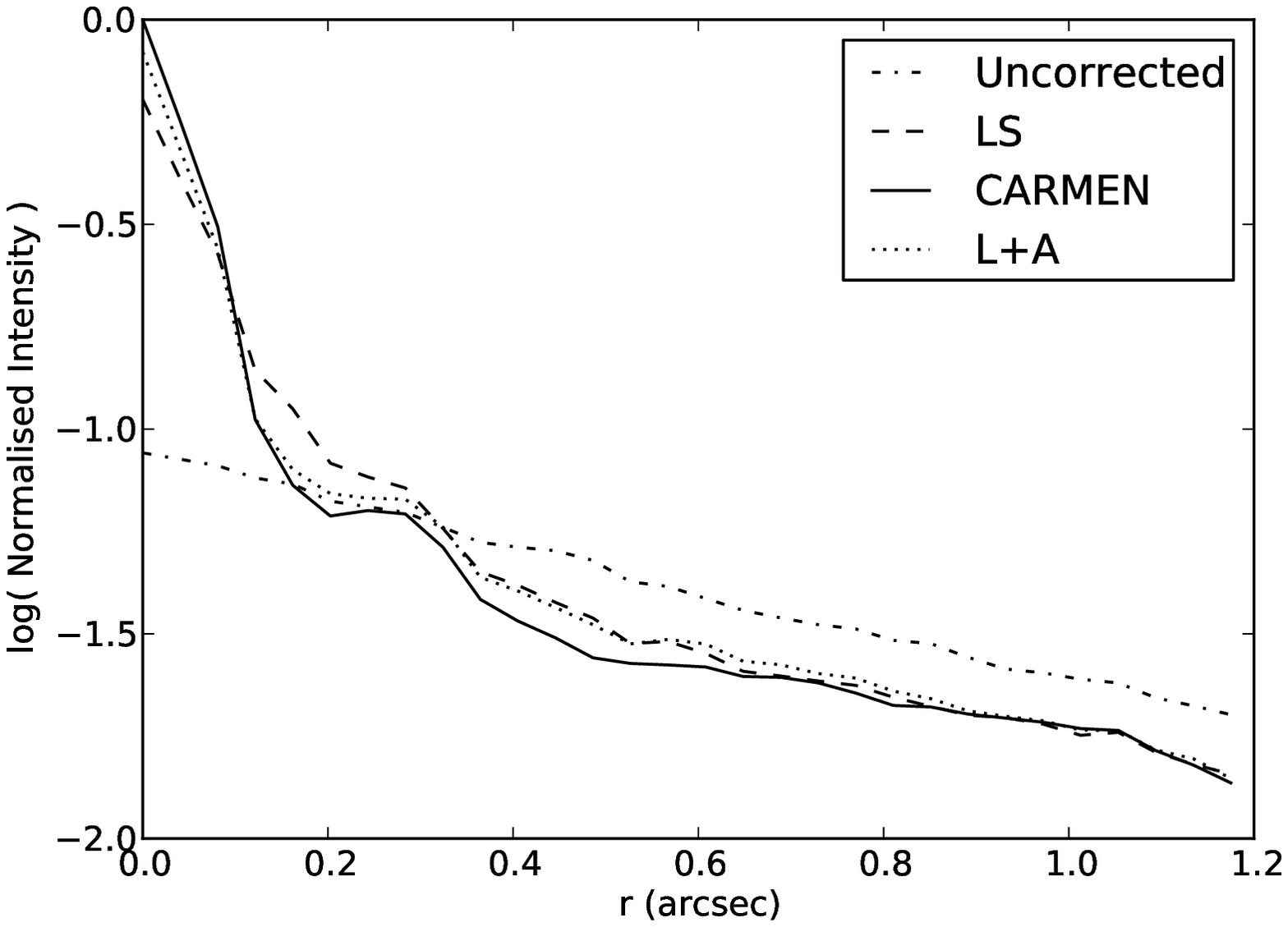}&
    \includegraphics[width=65mm]{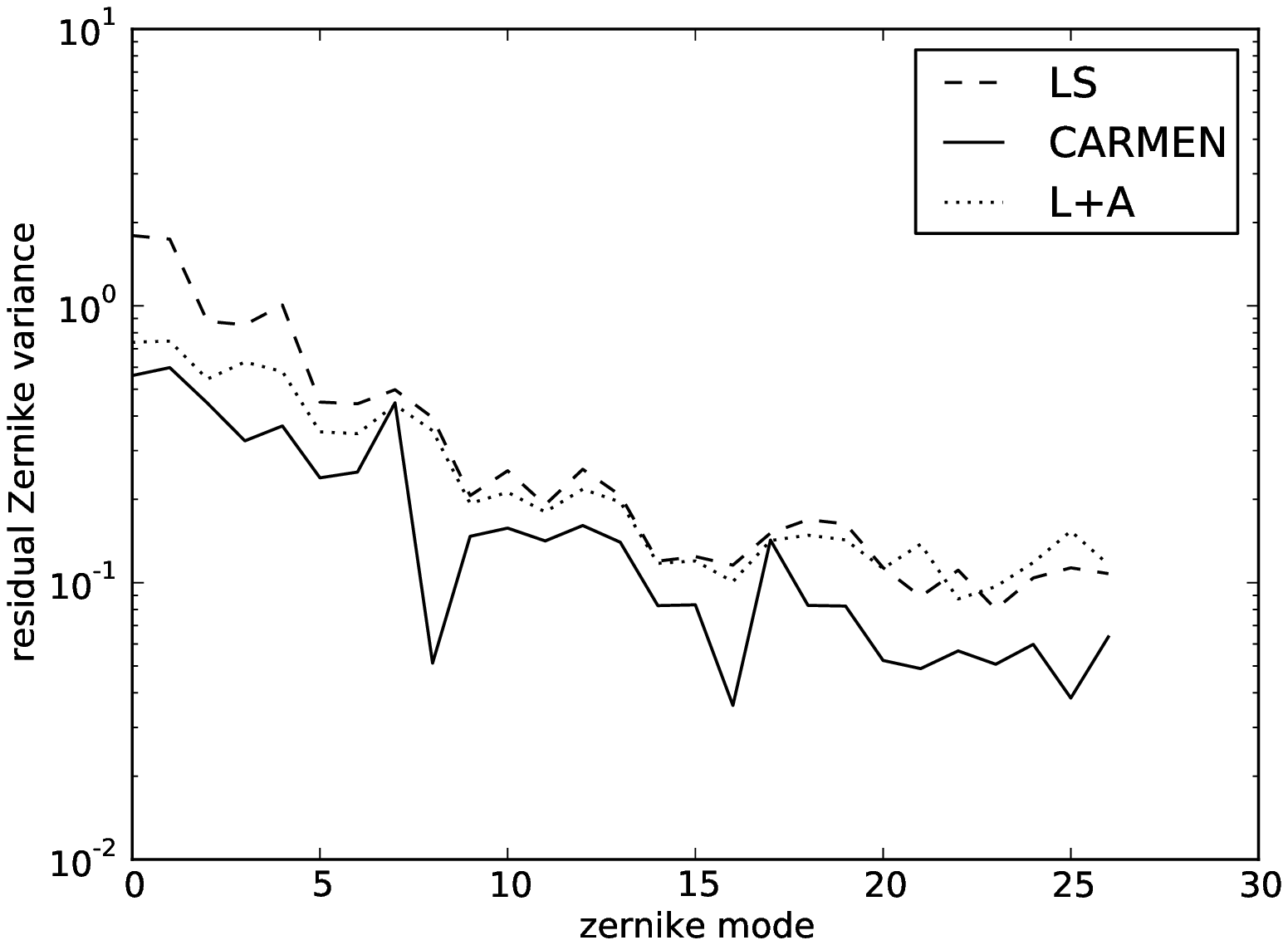}\\
           \mbox{\bf (a) } & \mbox{\bf (b)} 
      \end{array}$
   \caption{(a) Radial profiles of the simulated PSFs using the three test atmospheres ran sequentially to simulate a changing atmosphere. The residual WFE are 356~nm, 317~nm and 293~nm for LS, L+A and CARMEN reconstruction. (b) is the residual Zernike variance as a function of mode number.\comment{ We can see that CARMEN tomography, without any outside influence, is able to respond to the changing conditions during an observation.}}
   \label{fig:all_rad_prof}
\end{figure}

Figure~\ref{fig:all_rad_prof} (a) show the azimuthally averaged radial profiles for the scenario where each of the test cases are run sequentially. The LS and L+A were re-configured for each atmospheric test case. The WFE for the LS, L+A, and CARMEN tomographic techniques are 356~nm, 317~nm and 293~nm, this corresponds to Strehls of 0.198, 0.265 and 0.319, respectively. Figure~\ref{fig:all_rad_prof} (b) shows the residual Zernike variance on a mode by mode basis. We see that the residual variance is lower for CARMEN for every mode.

The test atmospheric profiles used above are all similar. We have also applied unrealistic extreme profiles to CARMEN to see if they will still be compensated. We introduce three more test cases, each with two turbulent layers and a 50\% split in turbulence strength, one at the ground and one at 5, 10 or 15~km. The LS and L+A were re-configured for each test and CARMEN was left unaltered. Table~3 shows the resulting metrics. The correction reduces with the altitude of the high turbulent layer because of the reduced fraction of overlap of the metapupils. We see that CARMEN functions with a wide range of altitudes for the high layer.
\begin{table*}   
\begin{minipage}{140mm}
\centering 
\begin{tabular}{| c|  c| c| c| c| c|}  
\hline
\hline
Reconstructor &  Altitude of high layer (m) & WFE (nm) & Strehl ratio\\ [0.5ex] 
\hline 
Uncorrected   &  5000  & 767 & 0.064 \\
LS                    &               & 293 & 0.289 \\
L+A                  &               & 269 & 0.353 \\
CARMEN        &              & 211 & 0.520\\  
\hline 
Uncorrected   &  10000 & 818 & 0.025\\  
LS                    &               & 465 & 0.066\\
L+A                  &               & 372 & 0.147\\
CARMEN        &               & 297 & 0.287\\ 
\hline 
Uncorrected   & 15000 & 815 & 0.026 \\  
LS                    &              & 574  & 0.043\\
L+A                  &              & 466 & 0.069 \\
CARMEN        &              & 390 & 0.127\\ 
\hline\hline
\end{tabular} 
\caption{Table of metrics for the three extreme test cases.}
\end{minipage}
\label{table:E}  
\end{table*}

\begin{figure}[htb]
   \centering
   \includegraphics[width=65mm]{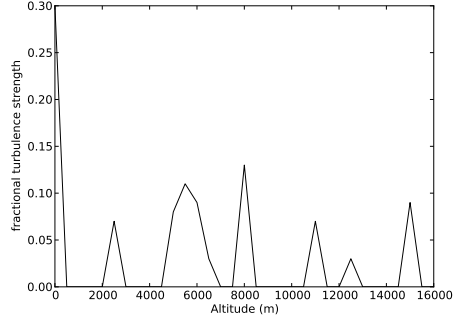}
   \caption{Arbitrary fictional seven layer turbulent profile used to test CARMENs ability to combine the response of many layers. The uncorrected WFE is 818~nm and the CARMEN residual WFE is 328~nm.}
   \label{fig:turb_profile}
\end{figure}

So far the test cases have involved small numbers of layers. Here we experiment with atmospheric profiles containing many layers. The residual WFE for a seven layer atmosphere (as shown in figure~\ref{fig:turb_profile}) with CARMEN was 328~nm compared to the uncorrected WFE of 818~nm. The integrated $r_{0}$ was 0.12~m. This shows that the network functions even with a large number of turbulent layers in the atmosphere without any modification or extra input.

\begin{figure}
   \centering
     $ \begin{array}{cc}
   \includegraphics[width=65mm]{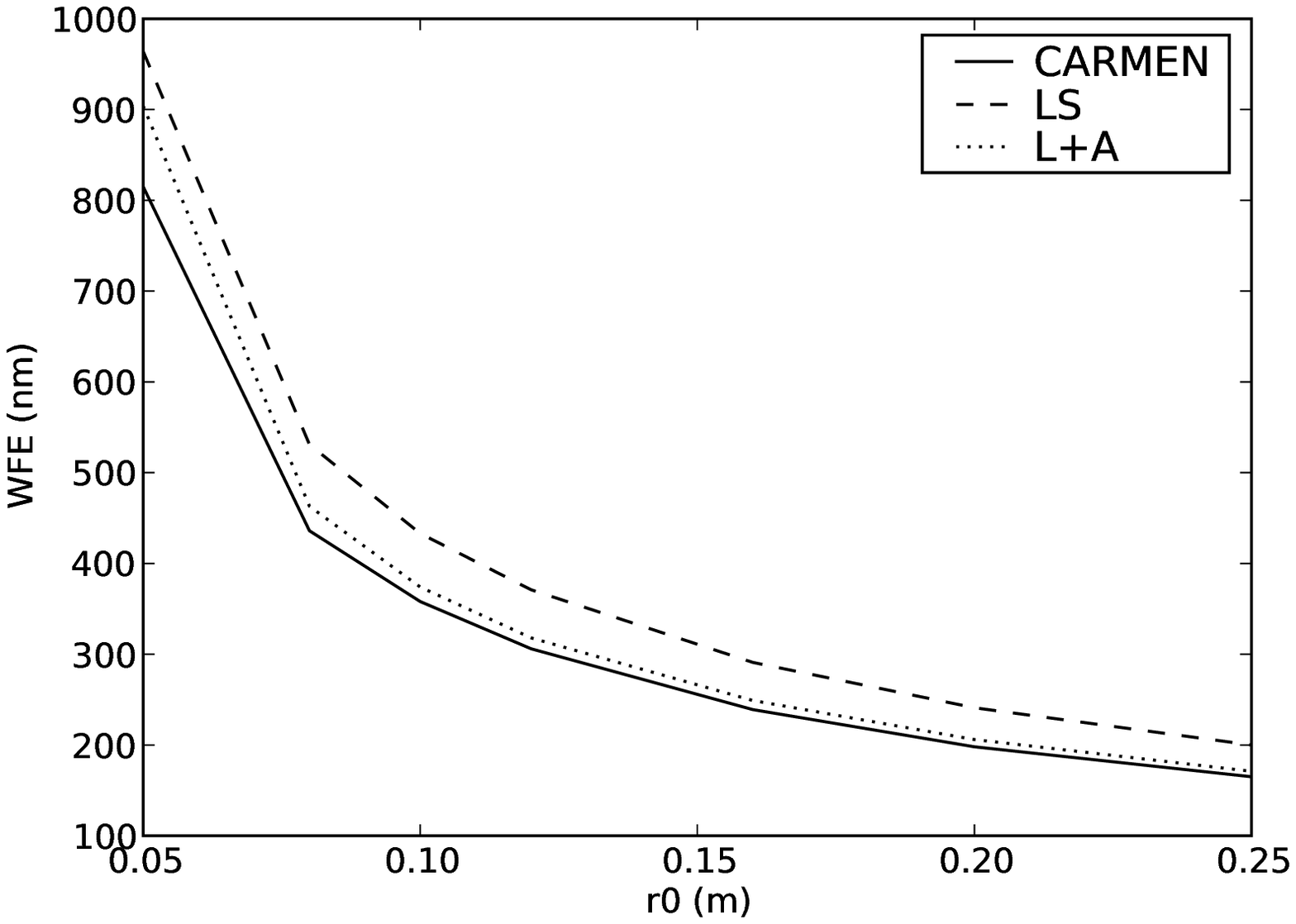}&
    \includegraphics[width=65mm]{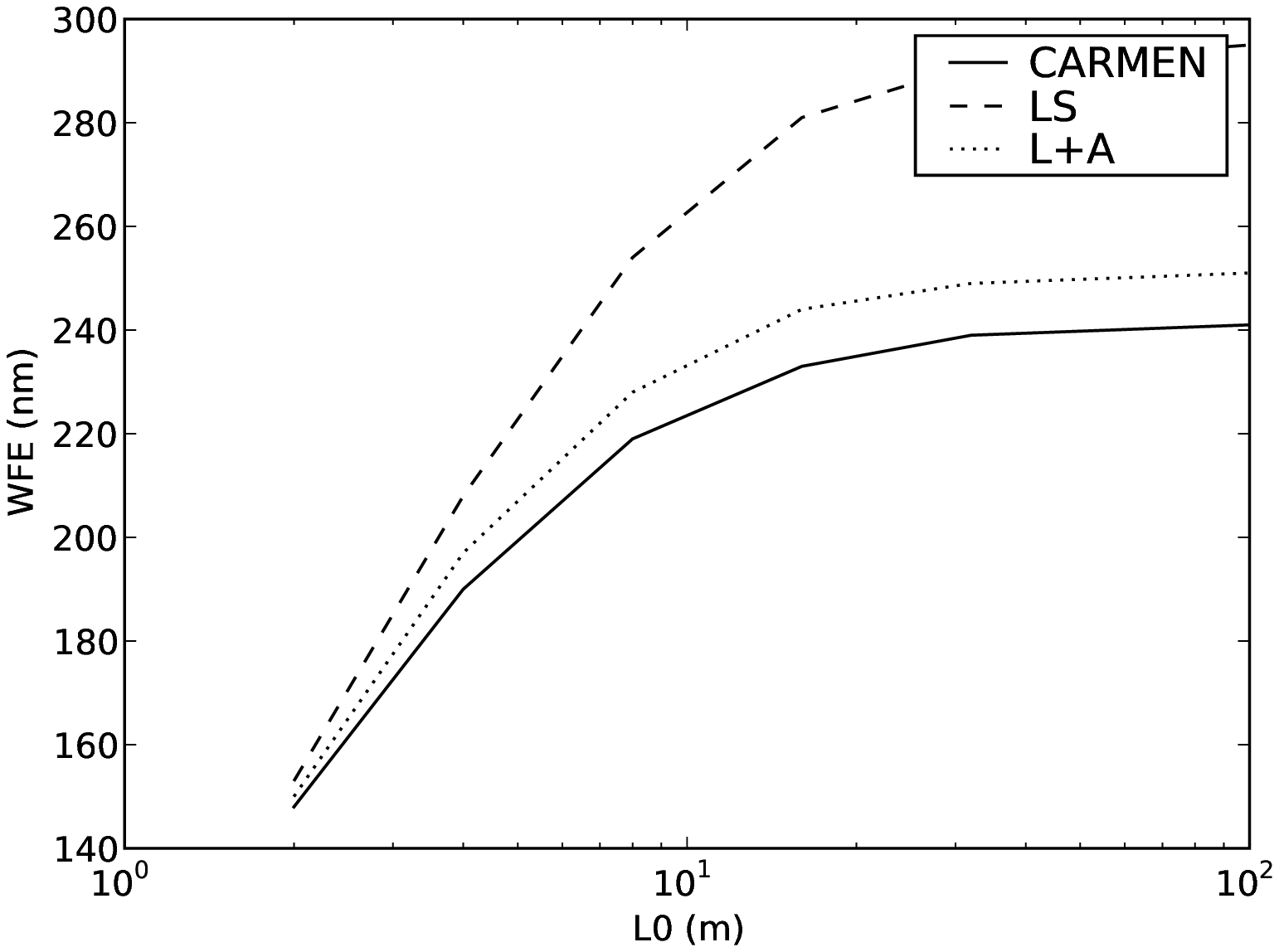}\\
   \mbox{\bf (a) } & \mbox{\bf (b)} 
   \end{array}$
   \caption{WFE as a function of integrated turbulence strength, $r_{0}$, (a) and of the outer scale, $L_{0}$, (b) using the atmospheric profile of test case 2.}
   \label{fig:r0}
\end{figure}

The plots in figure~\ref{fig:r0} show that CARMEN, although trained with a single value of the integrated turbulence strength, $r_{0}$, and outer scale, $L_{0}$, can actually correct for a wide range of realistic values. We varied $r_{0}$ between 0.05~m and 0.25~m and $L_{0}$ between 2~m ($D/2$, where $D$ is the diameter of the telescope) and 100~m ($\approx 25\times D$) and the observed pattern in correction is consistent with the other reconstructors.

\subsection{Simulation results with shot noise}

We have tested our reconstructor with simulated detector noise (shot noise and read noise) in the wavefront sensor. We assumed 100 photons per subaperture (which equates to an 11$^{\rm{th}}$ magnitude star and throughput of 50\% on a 4.2~m telescope), twenty by twenty pixels per subaperture and 0.2 electrons readout noise.

There are two approaches that we can take to train the ANN for noise. We can attempt to run the noisy WFS measurements through the original CARMEN trained without noise and we can try training a new ANN with slopes including centroid noise. After testing in simulation we find that the latter turns out to be a significantly better solution. Table~4 shows the resultant PSF metrics generated with reconstructors using WFS vectors including shot noise. We see that in the presence of shot noise the difference between CARMEN and the other reconstructors becomes even greater. This is expected as neural networks have been shown to be good at learning patterns in noisy data \cite{Tamura88}. The neural network is essentially de-prioritising higher order modes which are now indistinguishable from the noise. The noise was not included when training L+A and the conditioning parameter was altered to maximise the performance of the LS reconstructor.

\begin{table*}   
\begin{minipage}{140mm}
\centering 
\begin{tabular}{|c| c| c| c| c| c|}  
\hline
\hline
Test Name & Reconstructor &\multicolumn{4}{|c|}{Metrics} \\
\hline
& &Strehl ratio & FWHM (arcsec) & E50d (arcsec) & WFE (nm)\\ [0.5ex] 
\hline                 
atm 1 & Uncorrected & 0.048 & 0.319 & 0.482 & 643\\   
         & LS &  0.106  & 0.187 & 0.378 & 451\\
          &L+A   & 0.113 & 0.174 & 0.379 & 436\\
          & CARMEN  & 0.274 & 0.095 & 0.359 & 297 \\  [1ex]   
\hline    
atm 2 & Uncorrected & 0.025 & 0.458 & 0.633 & 817\\   
       & LS & 0.060 & 0.250 & 0.476 & 543\\            
 	& L+A   &0.055 & 0.254 & 0.524 & 547   \\       
          & CARMEN  & 0.158 & 0.105 & 0.477 & 368 \\  [1ex]   
\hline    
atm 3 & Uncorrected & 0.012 & 0.684 & 0.912 & 1087\\   
        & LS &0.021  & 0.455 & 0.771 & 756 \\
	& L+A & 0.020 & 0.455 & 0.773 & 751 \\
          & CARMEN  & 0.026 & 0.333 & 0.776 & 594 \\  [1ex]   
\hline\hline
\end{tabular} 
\caption{Table of PSF metrics for each tomographic reconstructor and test scenario including shot noise in the WFSs. All metrics, except WFE, are defined at 1650~nm.}
\end{minipage}
\label{table:results2}  
\end{table*}

\begin{figure}
   \centering
   $\begin{array}{cc}
       \includegraphics[width=65mm]{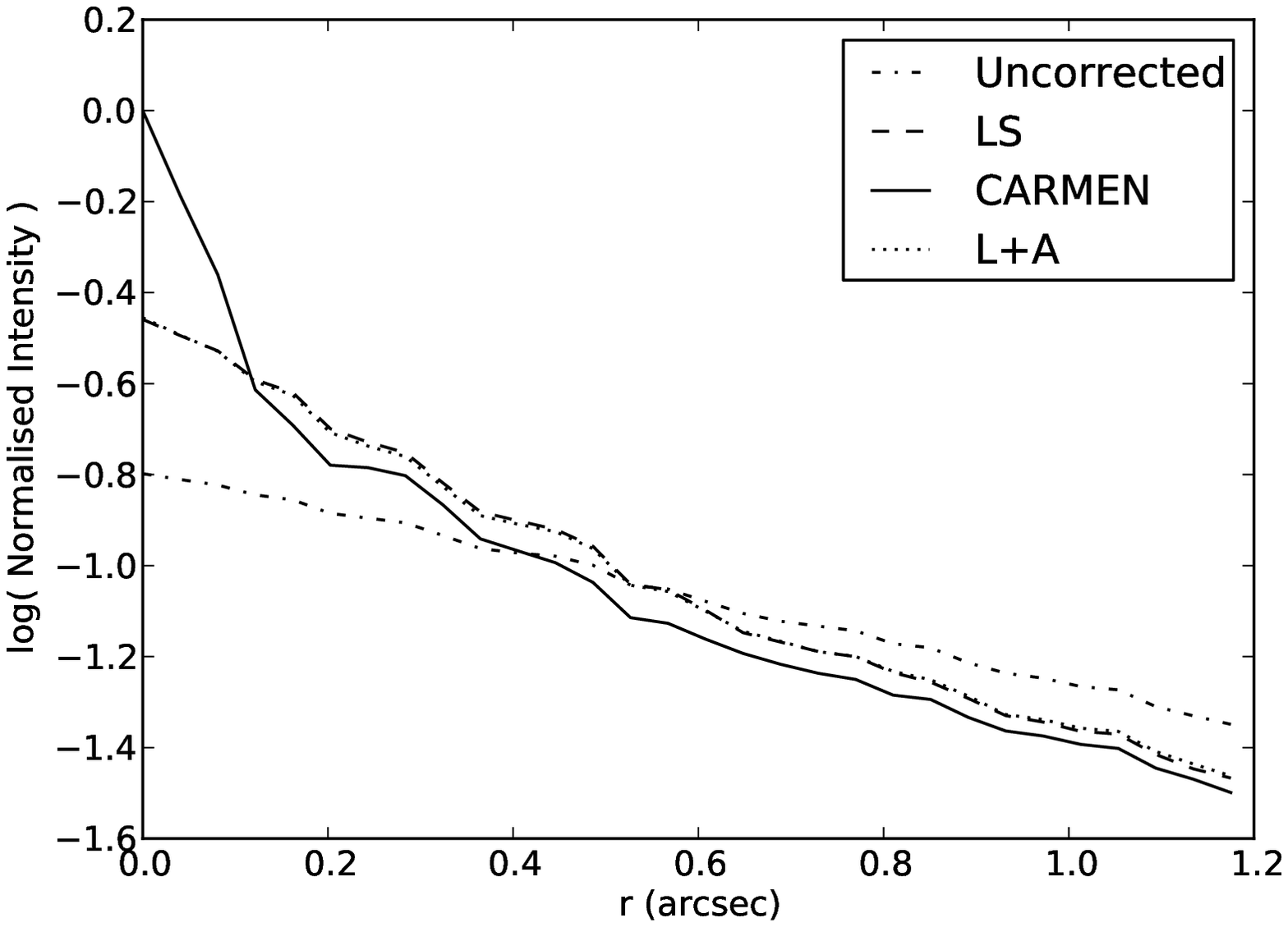}&
          \includegraphics[width=63mm]{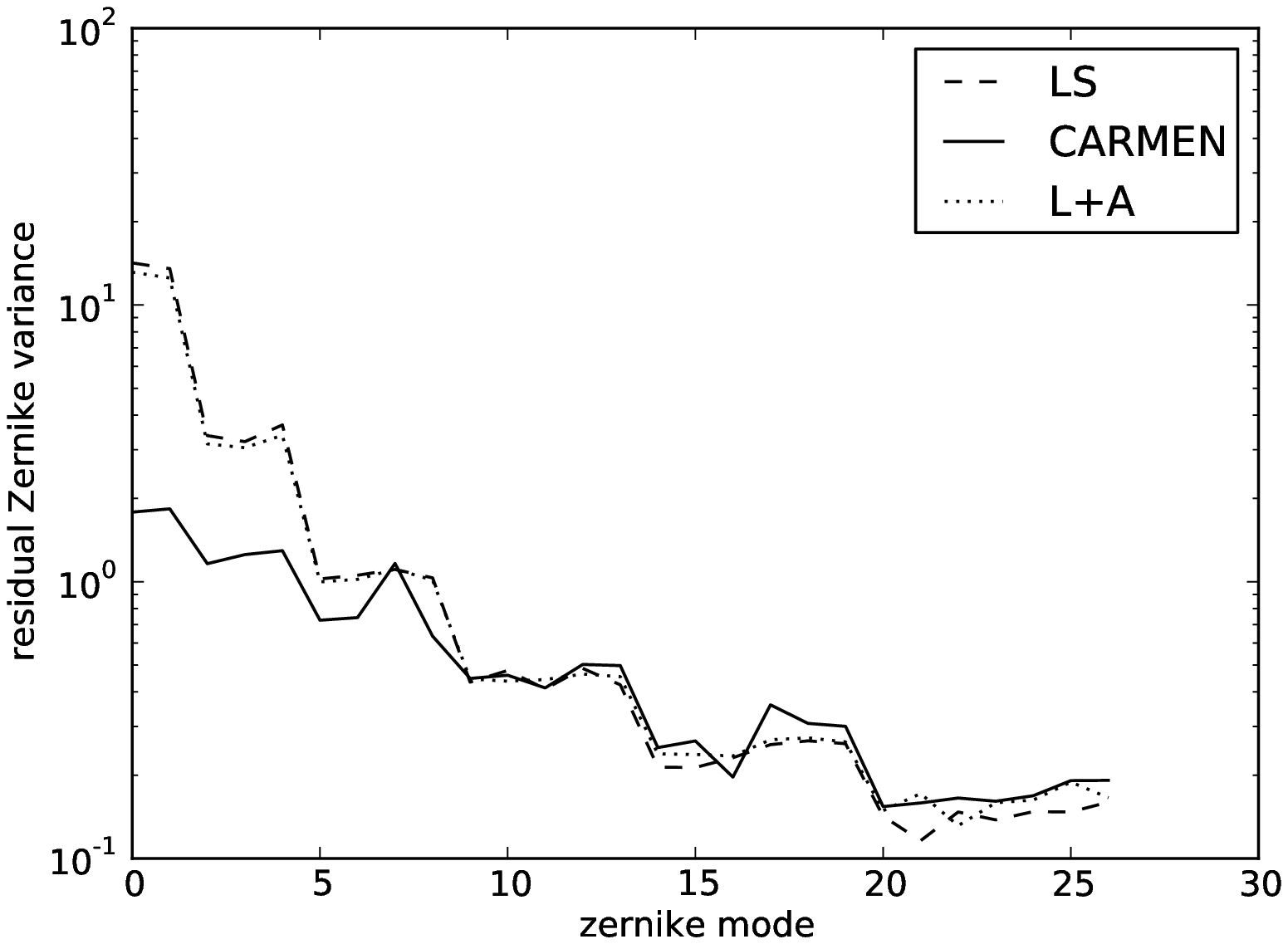}\\
             \mbox{\bf (a) } & \mbox{\bf (b)} 
   \end{array}$
   \caption{(a) Azimuthally averaged radial profiles of the uncorrected and LS, L+A and CARMEN reconstructed PSFs. Note that the LS and L+A radial profiles overlap almost perfectly. (b) Residual Zernike variance for the three reconstructors with WFS shot noise.}
   \label{fig:noisy}
\end{figure}

Figure~\ref{fig:noisy} (a) shows the radial profiles of the PSFs with the three different tomographic reconstructors with the median seeing atmospheric test case. The residual WFE for the uncorrected, LS, L+A and CARMEN reconstructors are 817, 543, 547 and 368~nm respectively. Figure~\ref{fig:noisy} (b) shows the variance of the residual Zernike coefficients ($\sum{(Z_{\mathrm{reconstructed}} - Z_{\mathrm{measured}})^2 / n}$, where $Z_{\mathrm{reconstructed}}$ are the reconstructed Zernike coefficients, $Z_{\mathrm{measured}}$ are the measured Zernike coefficients and $n$ is the number of iterations of the simulation) for each of the three reconstructors. We can see that CARMEN fits the low order modes better than the other methods. As most of the energy is concentrated in these modes this explains where the performance advantage of CARMEN comes from. However, in order to do this CARMEN must be trained with a dataset containing the same magnitude of shot noise.

\section{On-sky implementation}

In the work presented here we have only used natural guide stars. However, laser guide stars (LGS) will be required to increase the sky coverage. This will reduce the performance of any tomographic reconstructor due to the reduced overlap of the metapupils caused by focal anisplanatism of the beams \cite{Wilson96}. Although there are no problems with including LGSs in the training simulation there are other practical issues which may complicate an on-sky implementation. For example, we would need to train the network with the same Sodium column density profile and fratricide effects. Although ANNs have been shown to be robust the training simulation should incorporate all of these issues to optimise the performance.

So far all of the training has been done off-line in a simulation. This approach has the advantage that we can carefully select the training scenarios to optimise the performance. However, it might also be beneficial to have an additional on-line secondary correction which can tweak the output of CARMEN for the actual optical turbulence profile, WFS parameters, optical setup (e.g. misregistration errors) and centroid noise actually being observed and any other effect not included in the simulation. One option for this secondary correction would be to implement an additional neural network. As with the L+A technique an on-axis truth sensor would be required to train this network. Once trained this network will take in the vectors from the off-axis wavefront sensors and the output to the initial CARMEN prediction and output a new improved estimation of the on-axis phase aberrations. The disadvantage is that if we tune the tomographic reconstructor to the actual turbulence profile which then changes we will lose performance, as with the other reconstructors. This secondary network is currently under development and we plan to test it with on-sky data. 

\subsection{Extremely large telescopes}
An important question for AO instrument scientists is the scalability to ELT size telescopes. Due to the larger number of subapertures and guide stars involved, tomography on ELT scales becomes computationally more difficult. Although the training of the ANNs becomes exponentially more time consuming for larger telescopes (or more correctly, for larger number of subapertures) the computational complexity remains constant. Therefore,\comment{ given enough time} a network can be trained and implemented on ELT scale telescopes. Although we think that it might be possible to extrapolate the correction geometrically for any target direction it is worth noting that currently for every different asterism a new training is required. Therefore advanced planning is necessary. 

The strength of the ANN for tomographic reconstruction comes from its non-linear properties. The disadvantage of this means that the computational time required for each iteration scales badly in comparison to other linear techniques. However, the ANNs architecture and associated learning algorithms take advantage of the inherent parallelism in neural processing \cite{Hanggi00}. For specific applications such as tomographic reconstruction at ELT scales, which demand a high volume of adaptive real-time processing and learning of large data-sets in a reasonable time, the use of energy-efficient ANN hardware with truly parallel processing capabilities is more recomended \cite{Misra10}. A wide spectrum of technologies and architectures have been explored in the past. These include digital, analog, hybrid, FPGA based, and (non-electronic) optical implementations (\cite{Misra10} and references therein). Efficient neural-hardware designs are well known for achieving high speeds and low power dissipation. 

We are not able at this point to define the final computational necessities of an ANN tomographic reconstructor at ELT scales but, as an example of the capabilities of a neural-hardware implementation, a typical real-time image processing task may demand 10 teraflops, which is well beyond the current capacities of PCs or workstations today \cite{Misra10}. In such cases neurohardware appears an attractive choice and can provide a better cost-to-performance ratio even when compared to supercomputers.

\section{Conclusion}

We have presented and tested in simulation a novel and versatile tomographic reconstruction technique using an artificial neural network. We train the network with a number of simulated datasets designed to sample the full range of possible input signals. The data set comprises of a single turbulent layer positioned at a number of different altitudes in order to show the network as many different overlap fractions as possible. After testing several different training scenarios and network architectures we found that the simplest, with a single hidden layer, is the best. The reconstructor has been compared in simulation to a standard LS technique and to L+A.

We compare with LS and L+A only as a benchmark to show that the performance of CARMEN is on a par with other accepted reconstructor techniques. It is possible to optimise these reconstructors even more to obtain a better correction but we also believe that we can optimise CARMEN more by allowing the training process more time. Therefore, we do not want to draw conclusions about the magnitude of the correction at any one time. 

We have shown that the strength of CARMEN is two fold. By using an ANN we are able to train and apply a reconstructor which can adapt to wide range of atmospheric conditions. We tested CARMEN with the test case atmospheres used for the CANARY project and also with three more extreme atmospheres consisting of only two layers but with a 50\% split in the fractional turbulence strength and the high layer set to three different values separated by 5~km. We find that no change to CARMEN was needed even when the atmosphere changes drastically. We have also tested CARMEN with an atmosphere consisting of a large number of layers of varying fractional strength. Again we see that the ANN is able to successfully correct a large fraction of the turbulence induced phase aberrations. We varied the total integrated turbulence strength and the outer scale within a reasonable range of values (0.05~m $< r_{0} <$ 0.25~m, $D/2 < L_{0} < 25D$) and see that CARMEN follows the same trends as the two other reconstructors tested. If the atmosphere were to change dramatically during an observation the other reconstructors can be re-conditioned to deal with it. However, this does take time and is not required for the neural network approach.

The second strength of CARMEN is its ability to process shot noise corrupted centroid measurements. We have shown through Monte Carlo simulation that CARMEN is able to reconstruct the on-axis Zernike coefficients from noisy off-axis guide sources better than LS and L+A reconstructors. For example, using the CANARY median test case LS and L+A result in residual WFE of 543 and 547~nm respectively, CARMEN achieves a residual WFE of 368~nm. From analysis of the variance of the Zernike residuals we see that the majority of this improvement comes from the low order modes.

We have shown that, in simulation, ANNs can be used for tomographic reconstruction and can compete with other methods. The next step will be to test CARMEN in a more realistic situation. We intend to test CARMEN in the lab and later on-sky with CANARY.

\section*{Acknowledgements}
The author received a Postdoctoral fellowship from the School of Engineering at Pontificia Universidad Cat—lica de Chile as well as from the European Southern Observatory and the Government of Chile. D.~Guzman appreciates support from Pontificia Universidad Catolica, grant inicio No. 8/2010, TB acknowledges the Santander Mobility Grant. This work was partially supported by the Chilean Research Council grants Fondecyt-1095153 and Fondecyt-11110149 and by the Spanish Science and Innovation Ministry, project reference: PLAN NACIONAL AYA2010-18513. We would also like to thank Eric Gendron and Fabrice Vidal (LESIA) for their useful comments regarding the Learn and Apply method.

\end{document}